\documentclass[aps,prb,twocolumn,superscriptaddress]{revtex4-2}

\usepackage[dvipdfmx]{graphicx}% Include figure files
\usepackage{dcolumn}% Align table columns on decimal point
\usepackage{bm}% bold math
\usepackage{amsmath}
\usepackage{xcolor}

\begin{document}

%\preprint{}

\title{Low-energy photoexcitations inside the Mott gap in doped Hubbard and $t$-$J$ ladders}



\author{Sumal Chandra}
\email[]{sumalchandra@mbstu.ac.bd}
%\thanks{}
\affiliation{Department of Applied Physics, Tokyo University of Science, Katsushika, Tokyo 125-8585, Japan}
\affiliation{Department of Physics, Mawlana Bhashani Science and Technology University, Santosh, Tangail 1902, Bangladesh}

\author{Kazuya Shinjo}
%\email{kazuya.shinjo@riken.jp}
\affiliation{Computational Quantum Matter Research Team, RIKEN Center for Emergent Matter Science (CEMS), Wako, Saitama 351-0198, Japan}

\author{Shigetoshi Sota}
\affiliation{Computational Materials Science Research Team, RIKEN Center for Computational Science (R-CCS), Kobe, Hyogo 650-0047, Japan}
\affiliation{Quantum Computational Science Research Team, RIKEN Center for Quantum Computing (RQC), Wako, Saitama 351-0198, Japan}

\author{Seiji Yunoki}
\affiliation{Computational Quantum Matter Research Team, RIKEN Center for Emergent Matter Science (CEMS), Wako, Saitama 351-0198, Japan}
\affiliation{Computational Materials Science Research Team, RIKEN Center for Computational Science (R-CCS), Kobe, Hyogo 650-0047, Japan}
\affiliation{Quantum Computational Science Research Team, RIKEN Center for Quantum Computing (RQC), Wako, Saitama 351-0198, Japan}
\affiliation{Computational Condensed Matter Physics Laboratory, RIKEN Cluster for Pioneering Research (CPR), Saitama 351-0198, Japan}

\author{Takami Tohyama}
\email[]{tohyama@rs.tus.ac.jp}
%\thanks{}
\affiliation{Department of Applied Physics, Tokyo University of Science, Katsushika, Tokyo 125-8585, Japan}

%\date{\today}
\date{Received 27 September 2025; revised 28 November 2025; accepted 6 January 2026; published 15 January 2026}

\begin{abstract}
We investigate changes in the optical conductivity of doped Mott insulators by tuning ultrashort pump pulses to target either the Drude or low-energy absorption regions. Using a hole-doped two-leg Hubbard ladder and a four-leg $t$-$J$ ladder, we calculate the optical conductivity after pump by employing the time-dependent density matrix renormalization group. We find that a monocycle electric field pulse tuned to the Drude absorption reduces the Drude weight, accompanied by a slight enhancement in the mid-infrared (mid-IR) spectral weight. However, this enhancement diminishes as the pulse intensity increases. In contrast, a pump pulse tuned to the mid-IR absorption only affects the Drude weight. This behavior arises because the mid-IR absorption originates from magnetic excitations that do not couple directly to photons. These predictions can be tested experimentally by applying ultrashort low-energy pump pulses to cuprate materials.

\end{abstract}
\maketitle

%\twocolumn

\section{Introduction}
\label{Sec1}
The dynamics of holes doped into a Mott insulator induce both Drude and finite-frequency absorptions within the Mott gap in the optical conductivity~\cite{Dagotto1994}. In cuprate superconductors, a broad mid-infrared (mid-IR) absorption around 0.5~eV has been observed~\cite{Uchida1991}. The origin of the experimentally observed mid-IR absorption is attributed to a mixture of lattice and magnetic excitations~\cite{Mishchenko2008,Vidmar2009,Filippis2009}. Such broad mid-IR absorption is also present in two-leg ladder cuprates~\cite{Osafune1997}. The coexistence of both Drude and mid-IR absorptions is a hallmark of strongly correlated metallic systems.

Recent advances in laser technology have enabled the tuning of ultrashort pump pulses to match the Drude and mid-IR absorption regions using low-energy (mid-IR and THz) pulses. Indeed, mid-IR pulses have been applied to cuprate superconductors~\cite{Montanaro2024}. In the doped two-leg ladder Mott insulator Sr$_{14-x}$Ca$_x$Cu$_{24}$O$_{41}$, photo-pumping and pump-probe spectroscopy have been investigated~\cite{Fukaya2015,Hashimoto2016,Shao2019}, although the energy of the pump pulses used in these studies was near the edge of the Mott gap. Therefore, it is of great interest to understand how pump pulses tuned to the Drude and mid-IR absorptions affect the low-energy charge dynamics in strongly correlated metals. In particular, theoretical predictions regarding the differences between Drude pumping and mid-IR pumping will be valuable for guiding future experimental studies in these systems.

In this paper, we investigate the effects of photo-pumping on doped Mott insulators using few-cycle laser pulses with energies corresponding to the Drude and mid-IR absorption regions, and predict how the optical conductivity is modified by the pump. We employ a two-leg Hubbard ladder and a four-leg $t$-$J$ ladder as representative models of strongly correlated metals, and numerically examine the optical conductivity after pump using the time-dependent density matrix renormalization group (tDMRG) implemented with Legendre polynomials~\cite{Shinjo2022}. We find that a monocycle electric field pulse tuned to the Drude absorption reduces the Drude weight, accompanied by a slight enhancement of the mid-IR spectral weight. This enhancement diminishes as the pulse intensity increases. In contrast, pumping at the mid-IR absorption does not change the mid-IR weight but only reduces the Drude weight. This behavior is because mid-IR absorption originates from magnetic excitations, which do not couple directly to photons. These predictions can be tested experimentally by applying ultrashort low-energy pump pulses to cuprate materials.

This paper is organized as follows. In Sec.~\ref{Sec2}, we introduce the two-leg Hubbard ladder and the four-leg $t$-$J$ ladder, along with the method used to calculate the time-dependent optical conductivity. In Sec.~\ref{Sec3}, we investigate the photoinduced nonequilibrium properties of the two-leg Hubbard ladder after applying a monocycle pump pulse and a mid-IR pulse. In Sec.~\ref{Sec4}, we examine the four-leg $t$-$J$ ladder and discuss the behavior of the optical conductivity after the pump, with the aid of hole-density distributions and spin correlations.
Finally, we summarize our findings in Sec.~\ref{Sec5}. Throughout this paper, we set the speed of light $c$, the elementary charge $e$, the reduced Planck constant $\hbar$, and the lattice constant to unity.

\section{Model and method}
\label{Sec2}
We model a two-leg Hubbard ladder subjected to a spatially uniform time-dependent vector potential $A_\mathrm{leg}(t)$ along the leg direction as follows:
\begin{align}\label{Hub}
\mathcal{H}_\mathrm{Hub}=&-t_\mathrm{h} \sum_{i,\sigma} \Big( e^{iA_\mathrm{leg}(t)} \sum_\alpha c_{i,\alpha,\sigma}^\dag c_{i+1,\alpha,\sigma}  \nonumber \\
&+ c_{i,1,\sigma}^\dag c_{i,2,\sigma} + \mathrm{H.c.} \Big) + U\sum_{i,\alpha} n_{i,\alpha,\uparrow}n_{i,\alpha,\downarrow},
\end{align}
where $c_{i,\alpha,\sigma}^{\dagger}$ denotes the creation operator for an electron with spin $\sigma$ at rung $i$ on leg $\alpha = 1,2$, and $n_{i,\alpha,\sigma} = c_{i,\alpha,\sigma}^{\dagger} c_{i,\alpha,\sigma}$ is the corresponding number operator. The parameters $t_\mathrm{h}$ and $U$ represent the nearest-neighbor hopping amplitude and the on-site Coulomb repulsion, respectively. We consider a ladder consisting of $L = 16 \times 2$ lattice sites (16 rungs, $i=1,2,\cdots,16$), with open boundary conditions along both the leg and rung directions.

The four-leg $t$-$J$ ladder under the same spatially uniform vector potential is described by
\begin{align}\label{tJ}
\mathcal{H}_{tJ}=&-t_\mathrm{h} \sum_{i,\sigma} \Big( e^{iA_\mathrm{leg}(t)} \sum_\alpha \tilde{c}_{i,\alpha,\sigma}^\dag \tilde{c}_{i+1,\alpha,\sigma} \nonumber \\
&+ \sum_\alpha \tilde{c}_{i,\alpha,\sigma}^\dag \tilde{c}_{i,\alpha+1,\sigma} + \mathrm{H.c.} \Big) \nonumber \\
+& J\sum_{i,\alpha} \Big( \mathbf{S}_{i,\alpha}\cdot \mathbf{S}_{i+1,\alpha}-\frac{1}{4} n_{i,\alpha} n_{i+1,\alpha}  \nonumber \\
&+ \mathbf{S}_{i,\alpha}\cdot \mathbf{S}_{i,\alpha+1}-\frac{1}{4} n_{i,\alpha} n_{i,\alpha+1}  \Big),
\end{align}
where the operator $\tilde{c}_{i,\alpha,\sigma}^{\dagger} = c_{i,\alpha,\sigma}^{\dagger}(1 - n_{i,\alpha,-\sigma})$ creates an electron with spin $\sigma$ at rung $i$ on leg $\alpha=1,2,3,4$, enforcing the no-double-occupancy constraint. The parameter $J$ denotes the antiferromagnetic exchange interaction. We consider a $L=12 \times 4$ lattice (12 rungs, $i=1,2,\cdots,12$) with cylindrical geometry, where open boundary conditions are applied along the leg direction and periodic boundary conditions along the rung direction.

In both models, the electron density is set to 0.875, corresponding to a hole concentration of $x_\mathrm{h} = 1/8$ relative to half filling. The equilibrium properties of the two-leg Hubbard ladder and four-leg $t$-$J$ ladder at this doping level have been extensively studied using the density matrix renormalization group (DMRG)~\cite{Noack1997,Dolfi2015,Shen2023,Zhou2023,Tohyama1999,White1999,Scalapino2012,Dodaroo2017,Tohyama2018,Tohyama2020}. We adopt $U/t_\mathrm{h} = 10$ and $J/t_\mathrm{h} = 0.4$ as representative parameters for cuprate systems. Throughout this work, we use $t_\mathrm{h}$ and $t_\mathrm{h}^{-1}$ as units of energy and time, respectively. We note that $t_\mathrm{h}\sim 0.4$~eV in cuprates.

\begin{figure}[t]
\includegraphics[width=0.45\textwidth]{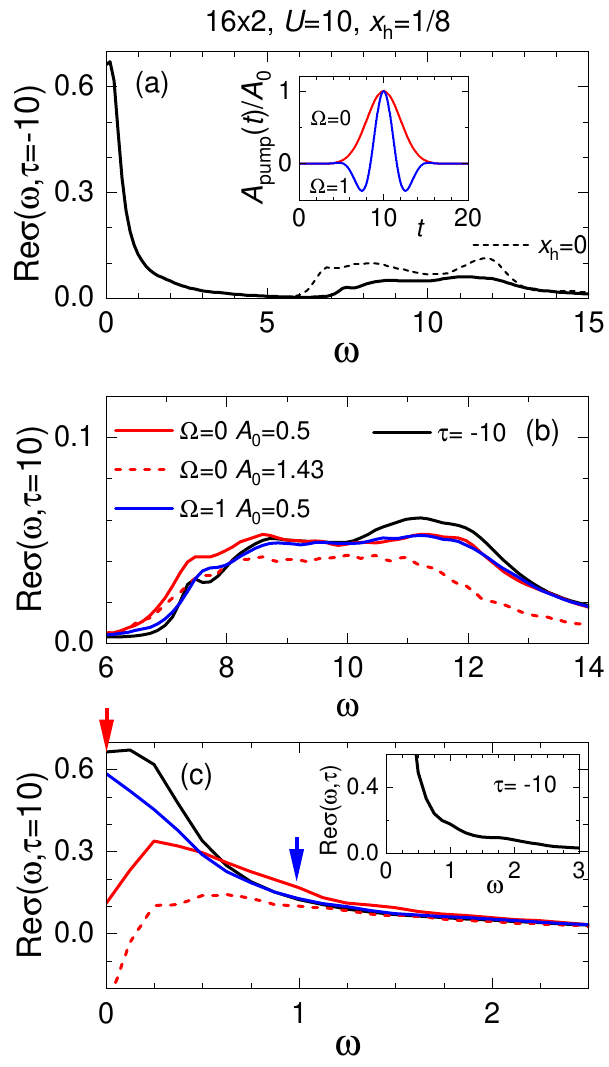}
\caption{$\mathrm{Re}\,\sigma(\omega,\tau)$ for the $x_\mathrm{h} = 1/8$ hole-doped two-leg Hubbard ladder of size $L = 16 \times 2$ with $U = 10$.  (a) Equilibrium case before pump at $\tau = -10$ (black solid line). For comparison, $\mathrm{Re}\,\sigma(\omega,\tau = -10)$ at half filling ($x_\mathrm{h} = 0$) is shown as the black dashed line.  Inset: Pump pulse $A_\mathrm{pump}(t)$ with $t_0 = 10$ and $t_\mathrm{d} = 2$ used in panels (b) and (c), where $\Omega = 0$ (red line) and $\Omega = 1$ (blue line).  (b) $\mathrm{Re}\,\sigma(\omega,\tau)$ above $\omega = 6$ and (c) below $\omega = 2.5$ at $\tau = 10$.  The red and blue solid lines correspond to $\Omega = 0$ and $\Omega = 1$, respectively, with $A_0 = 0.5$. The red dashed line represents the case of $\Omega = 0$ with $A_0 = 1.43$. The black solid line shows $\mathrm{Re}\,\sigma(\omega,\tau = -10)$, whose high-resolution behavior for the broadening factor of $\gamma=0.1$ is shown in the inset of (c), where small weight around $\omega\sim 1$ to 2 is evident.
In panel (c), the downward red and blue arrows indicate the position of pump frequency $\Omega$.
}
\label{Fig1}
\end{figure}

The pulse is assumed to have a time dependence given by $A_\mathrm{leg}(t)=A_\mathrm{pump}(t)+A_\mathrm{probe}(t)$ with 
\begin{equation}\label{Apump}
A_\mathrm{pump}(t)=A_0 e^{-(t-t_0)^2/(2t_\mathrm{d}^2)} \cos\left[\Omega(t-t_0)\right]
\end{equation}
for a pump pulse and 
\begin{equation}\label{Aprobe}
A_\mathrm{probe}(t)=A_0^\mathrm{pr} e^{-\left(t-t_0^\mathrm{pr}\right)^2/\left[2(t_\mathrm{d}^\mathrm{pr})^2\right]} \cos \left[\Omega^\mathrm{pr}(t-t_0^\mathrm{pr})\right]
\end{equation}
for a probe pulse, where both pulses have a Gaussian envelope for an oscillating part. We set the pump pulse parameters as $t_0 = 10$ and $t_\mathrm{d} = 2$ (see the inset of Fig.~\ref{Fig1}), and the pump frequency $\Omega$ is chosen to be $\Omega=0$, 1, and 2, which are below the energy of the Mott gap. For the probe pulse, we use $A_0^\mathrm{pr} = 0.001$, $\Omega^\mathrm{pr} = 10$, $t_\mathrm{d}^\mathrm{pr} = 0.02$, and $t_0^\mathrm{pr} = t_0 + \tau$, where $\tau$ denotes the delay time between the pump and probe pulses. Time-dependent wave functions $|\Psi(t)\rangle$ are computed from $t = 0$ to $80$ with a time step of $0.02$, under two conditions: with only $A_\mathrm{pump}(t)$, and with both $A_\mathrm{pump}(t)$ and $A_\mathrm{probe}(t)$. The simulations are performed using tDMRG based on the Legendre polynomial expansion~\cite{Shinjo2021b}. We retain $\chi = 2000$ density-matrix eigenstates, which ensures accurate evaluation of the optical conductivity discussed below.

The time-dependent current along the leg direction is calculated from the wave functions as
\begin{equation}\label{j}
j(t)=\frac{it_\mathrm{h}}{L}\sum_{i,\sigma} \left(e^{iA_\mathrm{leg}(t)} \sum_\alpha \left< c_{i,\alpha,\sigma}^\dag c_{i+1,\alpha,\sigma}\right> - \mathrm{H.c.}\right),
\end{equation}
where the expectation value $\langle \cdots \rangle = \langle \Psi(t) | \cdots | \Psi(t) \rangle$ is evaluated using the time-dependent wave function. For the $t$\nobreakdash-$J$ model, the operators $c_{i,\alpha,\sigma}^\dagger$ and $c_{i,\alpha,\sigma}$ are replaced by $\tilde{c}_{i,\alpha,\sigma}^\dagger$ and $\tilde{c}_{i,\alpha,\sigma}$, respectively. From the calculated current $j(t)$, we extract the contributions from the pump and probe pulses, denoted as $j_\mathrm{pump}(t)$ and $j_\mathrm{probe}(t,\tau)$, respectively: $j_\mathrm{pump}(t)$ is obtained by setting $A_0^\mathrm{pr} = 0$ in (\ref{Aprobe}), and $j_\mathrm{probe}(t,\tau)$ is defined as $j(t) - j_\mathrm{pump}(t)$ for a given $\tau$.

The nonequilibrium optical conductivity is defined as~\cite{Lu2015}
\begin{equation}\label{sigma}
\sigma(\omega,\tau) = \frac{1}{L}\frac{j_\mathrm{probe}(\omega,\tau) }{i(\omega +i\gamma)A_\mathrm{probe}(\omega)},
\end{equation}
where $A_{\mathrm{probe}}(\omega)$ and $j_\mathrm{probe}(\omega,\tau)$ denote the Fourier transforms of the probe vector potential $A_\mathrm{probe}(t)$ and the induced current $j_\mathrm{probe}(t,\tau)$ for $t > t_0^\mathrm{pr}$, respectively. The broadening factor $\gamma$ is set to 0.3 throughout this work, unless otherwise specified.
We calculate the static spin structure factor at momentum $\mathbf{q}=(q_\mathrm{leg},q_\mathrm{rung})$ for a $L=m\times n$ cylindrical cluster,
\begin{equation}\label{Szq}
S_z(\mathbf{q})=\left<S_{-\mathbf{q}}^zS_\mathbf{q}^z\right>
\end{equation} 
with
\begin{equation}\label{Sq}
S_\mathbf{q}^z=\sqrt{\frac{2}{(m+1)n}} \sum_{i,\alpha} \sin(q_\mathrm{leg}i) e^{-iq_\mathrm{rung}\alpha}S_{i,\alpha}^z \;,
\end{equation} 
where $q_\mathrm{leg}=n_\mathrm{leg}\pi/(m+1)$ ($n_\mathrm{leg}=1,2,\cdots,m$), $q_\mathrm{rung}=2n_\mathrm{rung}\pi/n$ ($n_\mathrm{rung}=0, \pm 1,\cdots,\pm(n/2-1), n/2$),  $S_{i,\alpha}^z$ is the $z$ component of $\mathbf{S}_{i,\alpha}$.

\section{Two-leg Hubbard model}
\label{Sec3}

Figure~\ref{Fig1}(a) shows the real part of the optical conductivity before pump, $\mathrm{Re}\,\sigma(\omega,\tau = -10)$, for the $x_\mathrm{h} = 1/8$ hole-doped two-leg Hubbard ladder with $L = 16 \times 2$ and $U = 10$~\cite{Shinjo2025}. For comparison, the result at half filling ($x_\mathrm{h} = 0$) is shown as a black dashed line, where absorption across the Mott gap appears above $\omega = 6$. Upon hole doping, the spectral weight above the Mott gap is reduced and shifts to higher energies, while a low-energy Drude-like excitation emerges. In the present $L = 16 \times 2$ system, a near-zero-energy excitation is observed, in contrast to smaller systems such as $L = 6 \times 2$~\cite{Fukaya2015,Hashimoto2016,Shao2019}, where the excitation emerges at finite energy due to finite-size effects. In addition to the Drude peak at $\omega = 0$, an additional small weight seems to exist around $\omega \sim 1$ to 2. The small weight becomes evident when a smaller value of $\gamma = 0.1$ is used, as shown in the inset of Fig.~\ref{Fig1}(c). This weight cannot be attributed to boundary effects from the open boundary condition along the leg direction, as a periodic $L = 12 \times 2$ system with $x_\mathrm{h} = 1/6$ also exhibits a similar feature in the same energy range (not shown). The additional weight corresponding to mid-IR absorption in cuprates is also evident in the four-leg $t$-$J$ model and will be discussed in Sec.~\ref{Sec4}. 

We apply pump pulses $A_\mathrm{pump}(t)$, as shown in the inset of Fig.~\ref{Fig1}(a), to the $L = 16 \times 2$ cluster. In the following, we consider three cases of pump pulses: $\Omega = 0$ and $\Omega = 1$ with $A_0 = 0.5$, and $\Omega = 0$ with $A_0 = 1.43$. The value $A_0 = 1.43$ is chosen such that the maximum amplitude of the electric field $E_\mathrm{pump}(t) = -\partial_t A_\mathrm{pump}(t)$ for $\Omega = 0$ matches that of the $\Omega = 1$ pulse with $A_0 = 0.5$. In other words, we prepare pump pulses with identical electric field amplitude but different $\Omega$ values, $\Omega=0$ and 1.

Figure~\ref{Fig1}(b) shows the high-energy part of $\mathrm{Re}\,\sigma(\omega,\tau)$ at $\tau = 10$, corresponding to $t_0^\mathrm{pr} = 20$. For the case of $\Omega = 0$ and $A_0 = 0.5$ (red solid line), a redshift of the remnant Mott gap is observed, similar to the response under a half-cycle pulse~\cite{Shinjo2025}. This redshift is attributed to the Stark effect induced by polarization, as in the half-cycle pulse case~\cite{Shinjo2025}. The polarization density at $\tau = 10$, when the probe pulse is applied, is defined as
\begin{equation}
p(t_0^\mathrm{pr}) = \int_0^{t_0^\mathrm{pr}} j_\mathrm{pump}(t)\,dt.
\end{equation}
We obtain $p(t_0^\mathrm{pr}) = 0.044$ for the $\Omega = 0$ pulse with $A_0 = 0.5$, and $p(t_0^\mathrm{pr}) = 0.026$ for the $\Omega = 1$ pulse with $A_0 = 0.5$. The difference in $p(t_0^\mathrm{pr})$ may contribute to a different amount of the redshift observed between the two cases [see red and blue lines in Fig.~\ref{Fig1}(b)]. In the case of $\Omega = 0$ with $A_0 = 1.43$, although the polarization $p(t_0^\mathrm{pr}) = -0.123$ is larger than in the other cases, the redshift is less pronounced. This may be due to significant spectral changes, particularly around $\omega \sim 12$, which affect the spectral weight near the edge of the remnant Mott gap. We note that the enhancement of reflectivity near the remnant Mott gap edge by a mid-IR pump pulse has been reported in a cuprate superconductor Bi$_2$Sr$_2$Ca$_{0.92}$Y$_{0.08}$Cu$_2$O$_{8+\delta}$~\cite{Montanaro2024}.

Figure~\ref{Fig1}(c) shows the low-energy part of $\mathrm{Re}\,\sigma(\omega,\tau = 10)$. In all three cases, the Drude peak near $\omega = 0$ (black line) is suppressed after pump. Among them, the suppression is weakest for $\Omega = 1$, and the spectral weight above $\omega = 0.7$ remains nearly unchanged. This indicates that pumping with $\Omega = 1$ induces minimal changes in the time-dependent wave function, despite the presence of a small absorption above $\omega = 1$ before pump, as discussed above. A similar behavior is found in the four-leg $t$-$J$ model (see Sec.~\ref{Sec4}).

For $\Omega = 0$ with $A_0 = 0.5$, the Drude weight is strongly suppressed, while spectral weight is enhanced around $\omega = 1$, consistent with reports under high-energy pumping with $\Omega = U$~\cite{Fukaya2015,Hashimoto2016,Shao2019} as well as under half-cycle pulses~\cite{Shinjo2025}. However, this enhancement disappears when the pump amplitude is increased to $A_0 = 1.43$, indicating that the spectral weight around $\omega = 1$ depends sensitively on the pulse amplitude. 

Furthermore, the low-energy spectral weight is strongly suppressed for the $A_0 = 1.43$ pulse. The strong suppression implies a significant reduction in the integrated $\mathrm{Re}\,\sigma(\omega,\tau = 10)$. Since the integrated weight is proportional to the absolute value of the kinetic energy along the leg direction~\cite{Shimizu2011}, this suggests that the time-dependent wave function exhibits a more localized character under a strong pump pulse. This behavior is qualitatively similar to that observed in the half-filled two-leg Hubbard ladder under a strong monocycle pulse~\cite{Tohyama2023}, where negative low-energy weight emerges inside the Mott gap.

The strong suppression of the weight is accompanied by a suppression of spectral weight just above $\omega = U$~\cite{Tohyama2023}. In the two-leg Hubbard ladder with $U = 10$ at half filling, a broad peak at $\omega = 12$ [see Fig.~\ref{Fig1}(a)], originating from localized doublon-holon excitonic states, is strongly suppressed~\cite{Tohyama2023}. Even in the 1/8 hole-doped case, the spectral weight at $\omega = 12$ is significantly reduced, as seen in the red dashed line of Fig.~\ref{Fig1}(b). Therefore, the appearance of strongly suppressed weight in Fig.~\ref{Fig1}(c) suggests that the same mechanism operative at half filling, i.e.,  photoinduced localization of doublon-holon pairs, also applies in the hole-doped regime under strong pump pulses, leading to an increased contribution of localized doublon-holon components in the time-dependent wave function~\cite{Tohyama2023}.

\begin{figure}[t]
\includegraphics[width=0.45\textwidth]{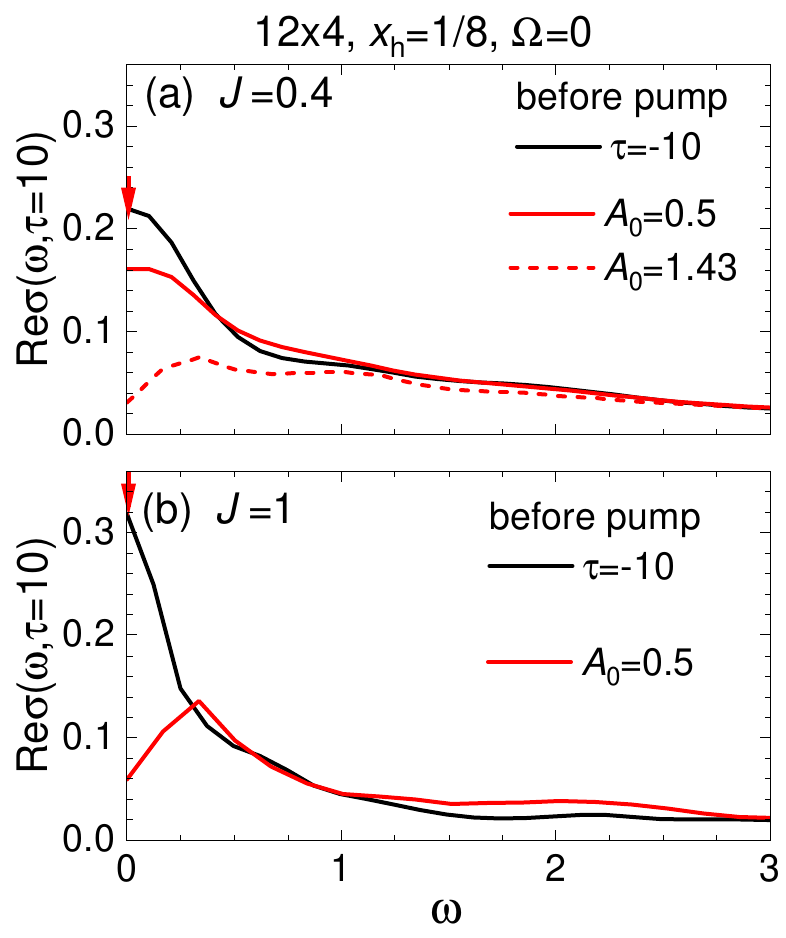}
\caption{Low-energy part of $\mathrm{Re}\,\sigma(\omega,\tau = 10)$ for the $x_\mathrm{h} = 1/8$ hole-doped four-leg $t$-$J$ ladder with $L = 12 \times 4$, pumped by the $\Omega = 0$ pulses. The equilibrium result at $\tau = -10$ is shown as the black line.  (a) $J = 0.4$: The red solid and red dashed lines correspond to $A_0 = 0.5$ and $A_0 = 1.43$, respectively.  (b) $J = 1$: The red line represents the case of $A_0 = 0.5$.  Downward red arrows indicate the position of pump frequency $\Omega$.}
\label{Fig2}
\end{figure}

\section{Four-leg $t$-$J$ ladder}
\label{Sec4}
Before pumping, $\mathrm{Re}\,\sigma(\omega,\tau = -10)$ (black line) for the $x_\mathrm{h} = 1/8$ hole-doped four-leg $t$-$J$ ladder with $L = 12 \times 4$ and $J = 0.4$ exhibits a Drude peak along with small humps around $\omega = 1$ and $2$, as shown in Fig.~\ref{Fig2}(a). These humps originate from string-type antiferromagnetic excitations, which are characteristic of two-dimensional Mott insulators~\cite{Stephan1990,Inoue1990,Dagotto1992,Jaklic2000,Tohyama2005,Shinjo2021a}. Experimentally, similar features have been observed in La$_{2-x}$Sr$_x$CuO$_4$~\cite{Uchida1991} and Sr$_{14-x}$Ca$_x$Cu$_{24}$O$_{41}$~\cite{Osafune1997} in the mid-IR region.  

Figure~\ref{Fig2} presents $\mathrm{Re}\,\sigma(\omega,\tau = 10)$ for $\Omega = 0$. In Fig.~\ref{Fig2}(a), corresponding to $J = 0.4$, the Drude weight below $\omega = 0.25$ is suppressed with increasing $A_0$. However, the low-energy spectral weight remains positive, in contrast to the two-leg Hubbard ladder, where strongly suppressed weight appears in the low-energy region. As discussed in Sec.~\ref{Sec3}, the suppression originates from the formation of localized holon-doublon pairs in the time-dependent wave function. In the present $t$-$J$ model, however, double occupancy is prohibited, and thus doublons are excluded. Consequently, localized holon-doublon pairs do not form, and the strong suppression of low-energy weight observed in the Hubbard model does not occur here. The suppression observed in the $t$-$J$ model is therefore attributed primarily to the increased kinetic energy of the existing hole carriers, resembling the effect of elevated temperature. This behavior has been referred to as the thermal effect~\cite{Shao2019}.

\begin{figure}[t]
\includegraphics[width=0.45\textwidth]{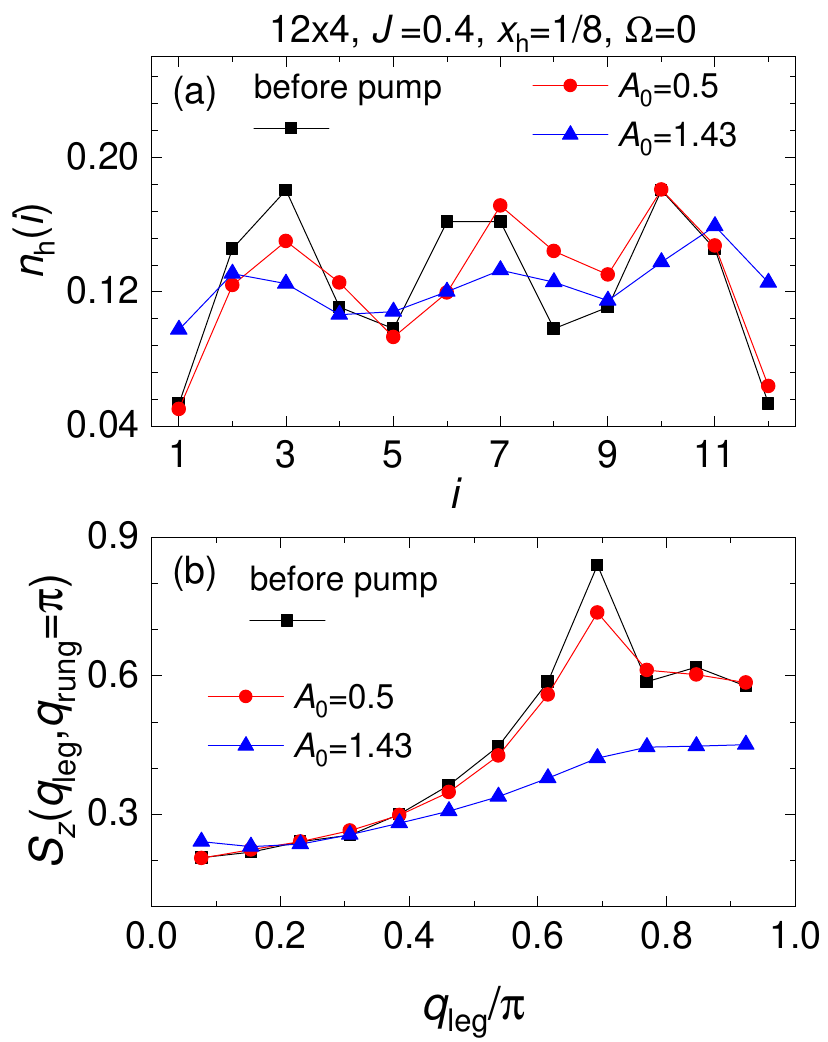}
\caption{Charge distribution and spin correlations in the $x_\mathrm{h} = 1/8$ hole-doped four-leg $t$-$J$ ladder with $L = 12 \times 4$ and $J = 0.4$.  (a) Hole density $n_\mathrm{h}(i)$ for each rung $i$.  (b) Static spin structure factor $S_z(q_\mathrm{leg}, q_\mathrm{rung} = \pi)$.  Red circles and blue upward triangles represent the results at $\tau = 10$ for the $\Omega = 0$ pulse with $A_0 = 0.5$ and $A_0 = 1.43$, respectively. The equilibrium case before pumping ($\tau = -10$) is shown by black squares. Solid lines are guides to the eye.}
\label{Fig3}
\end{figure}

To characterize the underlying electronic states after pump with $\Omega = 0$, we calculate the site-resolved hole density $n_\mathrm{h}(i)$ for each rung $i$. Before pump ($\tau = -10$), the holes exhibit a periodic modulation with a four-site periodicity (black squares), reminiscent of stripe-like charge order~\cite{Tranquada1995}. After pump ($\tau = 10$), this charge modulation becomes weaker and more uniform for $A_0 = 1.43$. Additionally, the hole distribution becomes asymmetric with respect to the center of the ladder ($i = 6$), reflecting the slow temporal variation of the electric field in the monocycle pulse. This asymmetry induces finite polarization, with $p(t_0^\mathrm{pr}) = 0.027$ for $A_0 = 0.5$ and $p(t_0^\mathrm{pr}) = 0.029$ for $A_0 = 1.43$.

Figure~\ref{Fig3}(b) shows the spin structure factor $S_z(\mathbf{q})$ at $q_\mathrm{rung} = \pi$. In equilibrium, a peak appears at $q_\mathrm{leg} = 0.69\pi$, which is close to the incommensurate antiferromagnetic wave vector $q_\mathrm{IC} = \pi(1 - 2x_\mathrm{h}) = 0.75\pi$ expected for stripe order~\cite{Tranquada1995}. After pumping with $A_0 = 0.5$, the peak position remains unchanged, although its intensity slightly decreases. In contrast, for $A_0 = 1.43$, the peak is smeared out and $S_z(\mathbf{q})$ becomes broader, indicating a weakening of antiferromagnetic correlations.

Returning to $\mathrm{Re}\,\sigma(\omega,\tau = 10)$, a small enhancement of absorption near $\omega = 1$, corresponding to the mid-IR region, is observed for $A_0 = 0.5$ in Fig.~\ref{Fig2}(a). This behavior is consistent with the result for the two-leg Hubbard ladder shown in Fig.~\ref{Fig1}(c). The enhancement in the mid-IR region is attributed to magnetic excitations activated by the pump pulse, with peak positions associated with string-type antiferromagnetic excitations~\cite{Shinjo2025}. For $A_0 = 0.5$, the pump pulse modifies the charge distribution while preserving magnetic correlations, as shown in Fig.~\ref{Fig3}. These subtle changes suppress the Drude weight and enhance the spectral intensity related to magnetic excitations.

In contrast, the enhancement at the mid-IR region in Fig.~\ref{Fig2}(a) diminishes with increasing $A_0$, accompanied by a weakening of antiferromagnetic correlations, as seen in Fig.~\ref{Fig3}(b). The magnetic origin of the mid-IR enhancement is further supported by the result for $J = 1$ in Fig.~\ref{Fig2}(b), where $\mathrm{Re}\,\sigma(\omega,\tau = 10)$ exhibits a broad increase in spectral weight around $\omega = 2$. The difference in the enhanced energy regions between Figs.~\ref{Fig2}(a) and \ref{Fig2}(b) reflects the variation in $J$, reinforcing the magnetic nature of the mid-IR response. 

\begin{figure}[t]
\includegraphics[width=0.4\textwidth]{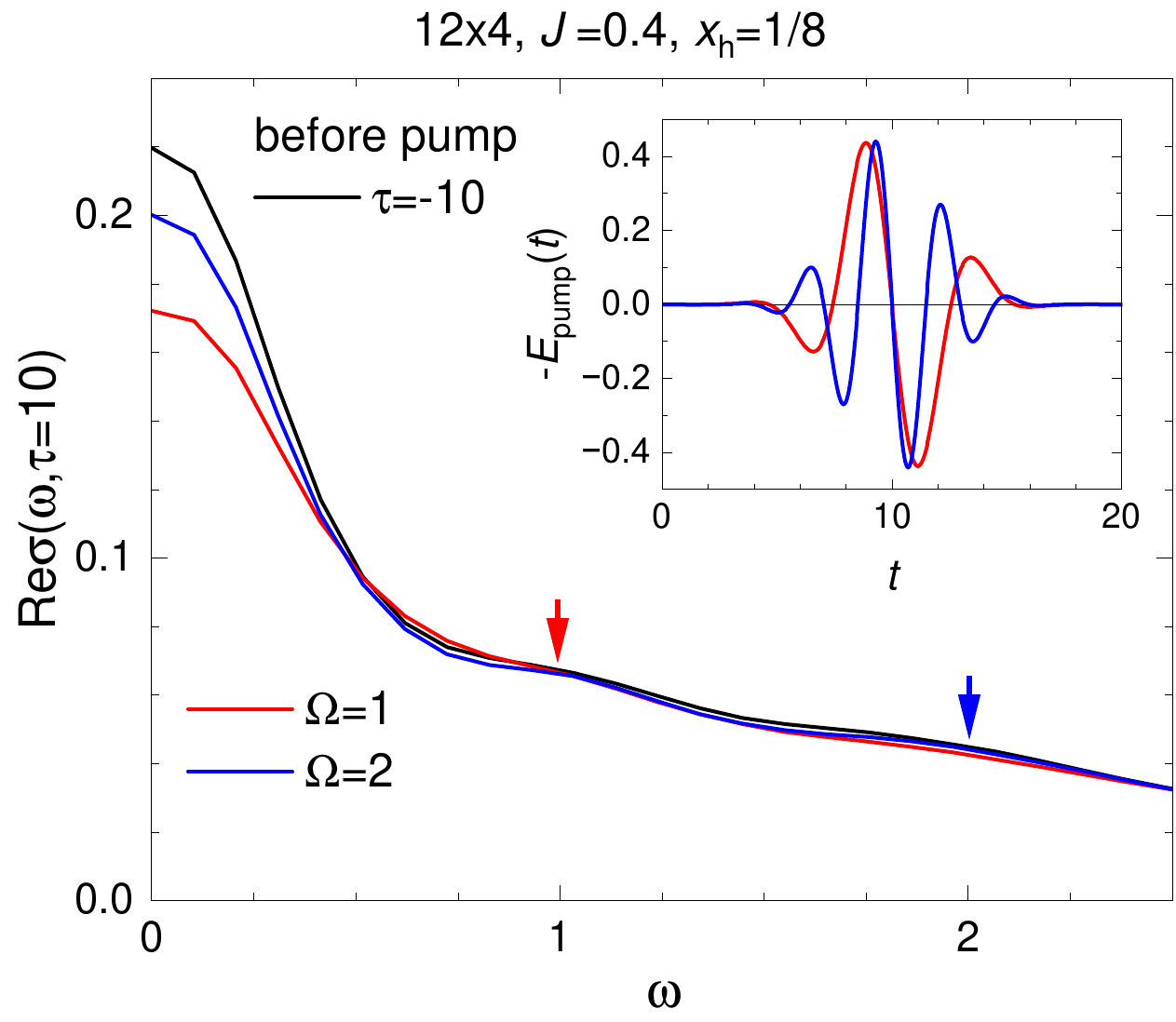}
\caption{Low-energy part of $\mathrm{Re}\,\sigma(\omega,\tau = 10)$ for the $x_\mathrm{h} = 1/8$ hole-doped four-leg $t$-$J$ ladder with $L = 12 \times 4$ and $J = 0.4$. The equilibrium result before pump ($\tau = -10$) is shown as the black line. The red and blue lines correspond to $\Omega = 1$ and $\Omega = 2$, respectively. The inset displays the electric field profile $E_\mathrm{pump}(t)$ for the pump pulse. The amplitude of the electric field is identical for all values of $\Omega$. Downward red and blue arrows indicate the position of pump frequency $\Omega$.}
\label{Fig4}
\end{figure}

Figure~\ref{Fig4} shows $\mathrm{Re}\,\sigma(\omega,\tau = 10)$ for pump pulses with $\Omega = 1$ and $\Omega = 2$. The amplitude $A_0$ for the $\Omega = 2$ pulse is adjusted such that the peak amplitude of the electric field $E_\mathrm{pump}(t)$ matches that of the $\Omega = 1$ pulse with $A_0 = 0.5$, resulting in $A_0 = 0.234$ for $\Omega = 2$. The corresponding electric field profiles are shown in the inset of Fig.~\ref{Fig4}. Application of the pump pulses leads to a slight suppression of the Drude weight. The degree of suppression correlates with the absorption intensity at the pump frequency prior to excitation. That is, the stronger the absorption at $\omega = \Omega$, the greater the reduction in metallic conduction. This behavior reflects an increase in kinetic energy induced by the pump pulse, analogous to thermal excitation, and is referred to as the thermal effect~\cite{Shao2019}.

\begin{figure}[t]
\includegraphics[width=0.45\textwidth]{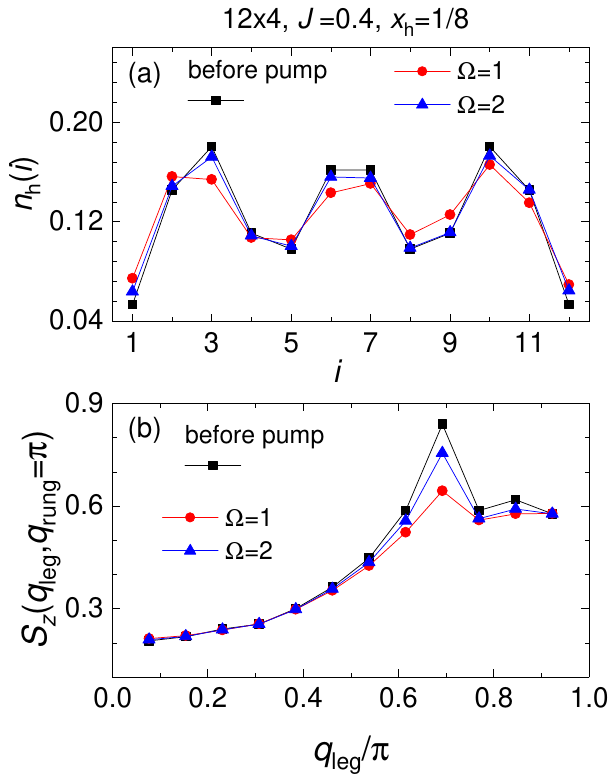}
\caption{Charge distribution and spin correlations in the $x_\mathrm{h} = 1/8$ hole-doped four-leg $t$-$J$ ladder with $L = 12 \times 4$ and $J = 0.4$.  (a) Hole density $n_\mathrm{h}(i)$ for each rung $i$.  (b) Spin structure factor $S_z(q_\mathrm{leg}, q_\mathrm{rung} = \pi)$.  Red circles and blue upward triangles represent the results at $\tau = 10$ for pump pulses with $\Omega = 1$ and $\Omega = 2$, respectively, under identical electric field amplitudes (see the inset of Fig.~\ref{Fig4}). The equilibrium case before pump ($\tau = -10$) is shown by black squares. Solid lines are guides to the eye.}
\label{Fig5}
\end{figure}

As shown in Fig.~\ref{Fig5}, the slight suppression of the Drude weight is accompanied by minor changes in the hole density and spin correlations. The four-site periodic charge distribution observed before pump is preserved for both $\Omega = 1$ and $\Omega = 2$ cases. Similarly, the peak in $S_z(q_\mathrm{leg}, q_\mathrm{rung} = \pi)$ at $q_\mathrm{leg} = 0.68\pi$ remains unchanged. These subtle modifications in charge and spin configurations primarily affect the Drude weight, while the spectral weight associated with string-type antiferromagnetic excitations remains unaffected.

\section{Summary}
\label{Sec5}

In summary, we have investigated the real part of the optical conductivity, $\mathrm{Re}\sigma(\omega,\tau)$, in pulse-excited states of hole-doped two-leg Hubbard and four-leg $t$–$J$ ladders using tDMRG. Few-cycle pump pulses tuned to the energies of Drude and mid-IR absorptions were applied, and the resulting optical responses were computed. A monocycle electric field pulse resonant with the Drude absorption reduces the Drude weight and slightly enhances the mid-IR component, although this enhancement diminishes with increasing pulse intensity. In contrast, excitation at the mid-IR absorption leads to a reduction in the Drude weight only. This behavior arises because mid-IR absorption originates from magnetic excitations that are not directly coupled to photons. These findings suggest that similar responses may be observed in cuprates under ultrashort, low-energy pump pulses.

\begin{acknowledgments}
This work was supported by the Japan Society for the Promotion of Science, KAKENHI (Grant Nos. JP23K13066, 	24K00560, 24K02948, and 25H01248) from the Ministry of Education, Culture, Sports, Science, and Technology, Japan. Numerical calculation was carried out using computational resources of HOKUSAI at RIKEN Advanced Institute for Computational Science, the facilities of the Supercomputer Center at Institute for Solid State Physics, the University of Tokyo, and supercomputer Fugaku provided by the RIKEN Center for Computational Science through the HPCI System Research Project (Project ID: hp240038).
\end{acknowledgments}

%\nocite{*}

%\begin{references}

\end{document}